\documentclass[twocolumn,showpacs,preprintnumbers,amsmath,amssymb]{revtex4}

\usepackage{graphicx}
\usepackage{revsymb}
\usepackage{dcolumn}
\usepackage{bm}

\begin{document}
\title{Control of mixed-state quantum systems by a train of short pulses}

\author{D. Sugny$^1$}
\email{dominique.sugny@u-bourgogne.fr}
\author{A. Keller$^2$}
\author{O. Atabek$^2$}
\author{D. Daems$^3$}
\author{C. M. Dion$^4$}
\author{S. Gu\'erin$^1$}
\author{H. R. Jauslin$^1$}
\affiliation{$^1$ Laboratoire de Physique de l'Universit\'e de Bourgogne, UMR CNRS 5027, BP 47870, 21078 Dijon, France}
\affiliation{$^2$ Laboratoire de Photophysique Mol\'{e}culaire du CNRS, Universit\'{e} Paris-Sud, B\^{a}t. 210 - Campus d'Orsay, 91405 Orsay Cedex, France}
\affiliation{$^3$ Center for Nonlinear Phenomena and Complex Systems, Universit\'{e} Libre de Bruxelles, 1050 Brussels, Belgium}
\affiliation{$^4$ Department of Physics, Ume\aa\ University, SE-90187 Ume\aa, Sweden}
\begin{abstract}
A density matrix approach is developped for the control of a mixed-state quantum system using a time-dependent external field such as a train of pulses. This leads to the definition of a target density matrix constructed in a reduced Hilbert space as a specific combination of the eigenvectors of a given observable through weighting factors related with the initial statistics of the system. A train of pulses is considered as a possible strategy to reach this target. An illustration is given by considering the laser control of molecular alignment / orientation in thermal equilibrium.
\end{abstract}
\maketitle

\section{Introduction}
The control of molecular dynamics by laser pulses has been a long-standing goal in the field of molecular physics, both from experimental and theoretical points of view (for a recent review, see \cite{seideman} and references therein). Molecular alignment or orientation induced by an intense laser field is one of the most challenging processes, with applications extending from chemical reactivity to nanoscale design \cite{seideman,warrer,dion,seideman1,brooks,aoiz}. In this context, many efforts, based on physical intuition or on optimal control strategies, have been made to construct control schemes, but with a limited success. For this reason, several works \cite{rama,girardeau,girardeau1,fu} have pointed out the connection between these results and tools of control theory. In particular, they have established kinematical bounds and determined the controllability of both pure and mixed-state molecular systems. More precisely, since the time-evolution operator is a unitary operator, only target states (corresponding to a pure or a mixed-state) which are unitarily equivalent to the initial state are kinematically attainable. This constraint is crucial since it determines the upper and lower bounds of the time-evolution of the expectation value of an arbitrary observable. However, it is important to remark that kinematical attainability is not synonymous with dynamical realizability, i.e. the possibility of finding a path of unitary operators, which satisfy the time-dependent Schr$\textrm{\"o}$dinger equation and bring the initial state to the target state within a class of physically realizable interactions. For control-linear processes (i.e. for Hamiltonians depending in a linear way on the control functions) of non-dissipative finite-level quantum systems, the dynamical realizability of the process depends on the dimension of the Lie algebra generated by the unperturbed Hamiltonian and the coupling operator \cite{rama,fu}.   

We have recently proposed a laser control strategy \cite{sugny,sugny3}, for a molecular system initially in a pure quantum state, aiming at the maximalization (or minimalization) of the expectation value of a given observable, which does not commute with the field-free Hamiltonian. This control scheme consists in applying sudden pulses each time a given quantity (such as the expectation value of the observable under consideration) reaches its maximum. We notice that there are several other methods for controlling molecular dynamics using, for instance, adiabatic passage techniques \cite{vitanov,guerin,guerin0}, factorizations of unitary operators \cite{schirmer} or optimal control schemes \cite{dion,zhu}. The main purpose of the present paper is to extend our control scheme to mixed-state quantum systems. Our concern is twofold : the identification of a well defined target density operator and the determination of a control scheme for reaching this target. Both steps are very general in their principle and their usefulness, such that the overall procedure may be extended to different control objectives. Alignment / orientation dynamics of a diatomic molecule are taken as examples, emphasizing the effects of non-zero temperature \cite{dion,benhaj,machholm,machholm1,ortigoso}. The drastic decrease of alignment / orientation with temperature basically results from an initial superposition of rotational states with $m\neq 0$ ($m$ being the projection of the total molecular angular momentum on the field polarization axis), which tends to misalign / misorient the molecule.

The identification of a target state as presented in this paper, together with the control strategy to reach it, is actually a very powerful tool. The analysis of the target state by itself may be also very instructive. An interesting counterintuitive example is the alignment / orientation of a diatomic molecule using differently polarized lasers. The simple examination of the two target states attainable by linear (no modification of $m$) or elliptical (inducing modifications of $m$) polarized lasers shows that they lead to alignment / orientation  properties that are very close to each other, rendering the use of elliptical polarizability unnecessary for diatomic systems. As for the control strategy, for non-dissipative systems (considered in this paper) the perturbations, described by unitary operators applied at given time intervals, yield spectacular results : the target state may be quite accurately and robustly reached within only a few perturbations. 

The paper is organized as follows : we outline the principles of the control strategy in Sec. \ref{section2}. Section \ref{section3} is devoted to the control of the alignment / orientation processes, taken on a parallel footing. Concluding remarks and prospective views are presented in Sec. \ref{section4}. Some details of the discussion are reported in Appendixes \ref{appb} and \ref{appc}.
\section{Principles of the control scheme} \label{section2}
This section outlines the principles of our control scheme in terms of a target state and the strategy to reach it. This has roughly an overall similitude with the scheme followed for a pure quantum state \cite{sugny,sugny3}, but the two approaches have to be contrasted at the level of the target state. In the pure-state case, it is a specific linear combination of rotational states, constituting a wavepacket evolving through the Schr$\textrm{\"o}$dinger equation. In the mixed-state case, the target is a density operator whose evolution is determined by the Von Neumann equation. The specific properties of the target density matrix cannot be inferred from a Boltzmann superposition of wavefunctions, separately time-propagated, i.e. it cannot be obtained from the analysis of Refs \cite{sugny,sugny3}.  

More specifically, we consider a mixed-state quantum system whose elements, the density operators $\rho(t)$, act on the Hilbert space $\mathcal{H}$. A density operator can be written as :
\begin{equation} \label{math0}
\rho=\sum_k \omega_k|\psi_k\rangle\langle \psi_k| \ ,
\end{equation}
where the $w_k$'s are the eigenvalues of $\rho$ which fulfill the following conditions : $0\leq w_k\leq 1$ and $\sum_k \omega _k=1$. The $|\psi_k\rangle$'s are the normalized eigenstates of $\rho$. Using atomic units ($\hbar=1$), $\rho(t)$ evolves according to the Von Neumann equation :
\begin{equation} \label{math1}
\frac{d}{dt}\rho(t)=i[\rho(t),H(t)] \ ,
\end{equation}
where $H(t)$ is the Hamiltonian of the system. The control being exerted through the application of a time-dependent external field, $H(t)$ is taken of the form 
\begin{equation} \label{math2}
H(t)=H_0+v(t)H_I \ ,
\end{equation}
where $H_0$ and $H_I$ are, respectively, the field-free Hamiltonian and the coupling operator. $v(t)$ is a real control function that can be turned on or off at will. For a given self-adjoint operator $\mathcal{O}$, which does not commute with $H_0$, and an initial state $\rho_0$, the goal of the control is to maximize or minimize the expectation value 
\begin{equation} \label{math3}
\langle \mathcal{O}(t)\rangle=\textrm{Tr}[\mathcal{O}\rho(t)] \ .
\end{equation}
More precisely, the goal is to find a path of unitary operators $U(t)$, which satisfy the time-dependent Schr$\textrm{\"o}$dinger equation [with an Hamiltonian of the form (\ref{math2})] and bring $\rho_0$ to the target state $\rho_F$, for which $\langle\mathcal{O}\rangle$ is a maximum or a minimum.\\
The set up of the control strategy consists of four steps. 

The first step is the reduction of the original physical Hilbert space $\mathcal{H}$ (possibly infinite) to a finite $N$-dimensional subspace $\mathcal{H}^{(N)}$. From the physical point of view, this reduction can be qualitatively justified by the fact that moderate perturbations, i.e. a finite number of applied pulses with moderate amplitude, can only transfer finite amounts of energy to the system, which stays thus essentially confined in a finite-dimensional subspace \cite{sugny}. This reduction drastically simplifies the study of both kinematical constraints and dynamical realizability as we will show below. In the subspace $\mathcal{H}^{(N)}$, we consider the reduced operator $\mathcal{O}^{(N)}$, which is defined by 
\begin{equation} \label{math4}
\mathcal{O}^{(N)}=P^{(N)}\mathcal{O}P^{(N)} \ ,
\end{equation}
$P^{(N)}$ being the projector on $\mathcal{H}^{(N)}$. To simplify the notation, we omit the superscript $(N)$ for the operators $H_0$ and $H_I$ in the rest of the paper. Finally, we also assume that the field-free evolution is periodic with period $T_0$, which means that all the fundamental frequencies of $H_0$ in $\mathcal{H}^{(N)}$ are commensurate and integer multiples of a fundamental frequency $\omega=2\pi/T_0$.

We next define the initial state. We assume that the system is initially in thermal equilibrium with a bath at temperature $T$. The initial density operator $\rho_0$ is then the canonical density operator, written here in the basis of the eigenvectors $|n\rangle$ of $H_0$ :
\begin{equation} \label{math5}
\rho_0=\frac{1}{Z}\exp\left[-\frac{H_0}{k_B T}\right]=\frac{1}{Z}\sum_{n=1}^N \exp\left[-\frac{E_n}{k_B T}\right] |n\rangle\langle n| \ ,
\end{equation}
where $Z=\sum_{n=1}^N \exp[\frac{-E_n}{k_B T}]$ is the partition function, $k_B$ the Boltzmann constant, and $H_0=\sum_{n=1}^N E_n|n\rangle\langle n|$. The difference of treatment between pure and mixed-state quantum systems at this first step is that, while the initial state of a pure quantum system is usually a single eigenstate of $H_0$  \cite{sugny3}, that of a mixed-state system is a density operator involving a thermal average of all eigenstates of $H_0$. 

The second step consists in determining a target state $\rho_F$ which both maximizes (or minimizes) the expectation value $\langle \mathcal{O}^{(N)} \rangle$ and is unitarily equivalent to the initial state. The general characterization of this target state is adapted from a recent work of Girardeau and co-workers \cite{girardeau,girardeau1}. It can be shown that the density operator which maximizes $\langle\mathcal{O}^{(N)}\rangle$ commutes with $\mathcal{O}^{(N)}$ \cite{girardeau1}. These two operators can therefore be simultaneously diagonalized. Moreover, due to the constraint of unitary evolution, $\rho_0$ and $\rho_{F}$ have the same eigenvalues $\omega_k$ with the same multiplicity.  The expectation value $\langle\mathcal{O}^{(N)}\rangle=\textrm{Tr}[\mathcal{O}^{(N)}\rho_{F}]$ can thus be written as follows :
\begin{equation} \label{math6}
\textrm{Tr}[\mathcal{O}^{(N)}\rho_F]=\sum_{k=1}^N \chi_k\omega_{\sigma_k} \ ,
\end{equation}
where the $\chi_k$'s are the eigenvalues of $\mathcal{O}^{(N)}$ and the $\omega_{\sigma_k}$'s the optimal permutation of the eigenvalues $\omega_k$'s of $\rho_0$, which is defined as the permutation which associates the highest eigenvalue of $\rho_F$ with the highest one of $\mathcal{O}^{(N)}$, the second highest with the second one, and so forth. We denote $\rho^{(N)}_F$ the resulting density operator which maximizes $\mathcal{O}^{(N)}$ and which is precisely our target in the reduced space $\mathcal{H}^{(N)}$. This density operator is not unique if the spectrum of $\mathcal{O}^{(N)}$ is degenerate. Here again, we emphasize that the target to be reached in our control scheme of mixed-state quantum systems is a specifically build density operator, as opposite to the case of a pure-state quantum system where the target corresponds to the eigenvector of $\mathcal{O}^{(N)}$ associated with the highest eigenvalue \cite{sugny3}.

The third step deals with the dynamical realizability of the target state. In other words, an initial density operator and a target being given, the question is to know if there exists a path of unitary operators that bring the system from one to the other. A sufficient condition to ensure the dynamical attainability of the target is the complete controllability of the system. We recall that a control-linear system is completely controllable if the dynamical Lie algebra $\mathcal{L}$ generated by the skew-hermitian operators $iH_0$ and $iH_1$ is $u(N)$. However, it is noted that this condition is a strong requirement which is not necessary. A less strong condition is the notion of connectivity where a non-zero amplitude of the evolution operator between the initial and the target states only is required \cite{wu}. Here, in order to clarify the discussion in the general context we assume the complete controllability of the process.

The last step of the control strategy consists in explicitely finding a control scheme which allows us to attain this target. We devise two control strategies which are in the spirit of our analysis for pure-state quantum systems \cite{sugny,sugny3,averbuch}, while they differ in the definition of the targets. More precisely, they consist respectively in applying pulses each time $\langle\mathcal{O}^{(N)}(t)\rangle$ (strategy $S1$) or $\langle \rho^{(N)}_F|\rho (t)\rangle=\textrm{Tr}[\rho^{(N)}_F \rho(t)]/\textrm{Tr}[(\rho^{(N)}_F)^2]$ (strategy $S2$) reaches a global maximum within a period $T_0$ of the field-free dynamics. The unitary operator $U(t)$ satisfying the Schr$\textrm{\"o}$dinger equation can then be written as a product of terms of the form :
\begin{equation} \label{math7}
\exp[-iH_0\Delta t]U_{\tilde A} \ ,
\end{equation}
the first factor $\exp[-iH_0\Delta t]$ corresponding to the field-free motion and the second one to the application of the pulse. $\Delta t$ is here the time delay between pulses and $\tilde{A}$ a real control parameter. Moreover, we assume that $U_{\tilde{A}}$ commutes with $\mathcal{O}^{(N)}$ or with $\rho^{(N)}_F$ (depending on the strategy $S1$ or $S2$) so that its application does not alter $\langle\mathcal{O}^{(N)}(t)\rangle$ or $\langle \rho_F|\rho (t)\rangle$. More precisely, denoting the operator $\mathcal{B}$ as $\mathcal{O}^{(N)}$ or $\rho_F^{(N)}$, we have 
\begin{equation} \label{math8}
\textrm{Tr}[\rho(t)\mathcal{B}]=\textrm{Tr}[U_{\tilde{A}}\rho(t)U_{\tilde{A}}^{-1}\mathcal{B}] \ ,
\end{equation}
which proves the claim. As opposed to the value of the function $\textrm{Tr}[\rho(t)\mathcal{B}]$, its derivative can be altered by the application of the pulse. Indeed, the slope given, before the excitation, by
\begin{equation} 
\frac{d}{dt}\textrm{Tr}[\rho(t)\mathcal{B}]|_{t_0-0}=\textrm{Tr}[\rho(t_0)[H_0,\mathcal{B}]]=0 \ ,
\end{equation}
changes into 
\begin{equation} \label{math14}
\frac{d}{dt}\textrm{Tr}[\rho(t)\mathcal{B}]|_{t_0+0}=\textrm{Tr}[\rho(t_0) U_{\tilde{A}}^{-1}[H_0,\mathcal{B}]U_{\tilde{A}}] \ ,
\end{equation}
when a pulse is applied at time $t_0$.

We next notice that the two functions $\langle\mathcal{O}^{(N)}(t)\rangle$ and $\langle \rho^{(N)}_F|\rho (t)\rangle$ are periodic, continuous and differentiable (under field-free evolution), leading thus to an increasing (possibly constant), bounded and therefore convergent sequence of maxima. The limit of this sequence is a fixed point, i.e. a density operator $\rho_f$ such that  
\begin{equation} \label{math9}
\textrm{Tr}[\rho_f\mathcal{B}]=\textrm{Tr}[U_{\tilde{A}}\rho_fU_{\tilde{A}}^{-1}\mathcal{B}] \ ,
\end{equation}
is a global maximum within a period $T_0$, for any value of $\tilde{A}$. In other words, for the mixed state $\rho_f$ this means that one cannot find a value of $\tilde{A}$ such that the dynamics passes through (under free evolution) a maximum strictly larger than the one just before the pulse. We denote by $\mathcal{F}$ the set of the fixed points. Due to the difficulty to determine $\mathcal{F}$ exactly, we consider a larger set $\mathcal{S}$, defined as the set of density operators $\rho$ which fulfill the following requirements :
\begin{equation} \label{math11}
\textrm{Tr}[\rho [H_0,\mathcal{B}]]=0 \ ,
\end{equation}
and
\begin{equation} \label{math12}
\textrm{Tr}[\rho U_{\tilde{A}}^{-1}[H_0,\mathcal{B}]U_{\tilde{A}}]=0 \ ,
\end{equation}
for all values of $\tilde{A}$. If the slope of $\textrm{Tr}[\rho(t_0)\mathcal{B}]$ at time $t_0$ undergoes a change from zero before the pulse :
\begin{equation} \label{math13}
\frac{d}{dt}\textrm{Tr}[\rho(t)\mathcal{B}]|_{t_0-0}=\textrm{Tr}[\rho(t_0)[H_0,\mathcal{B}]]=0 \ ,
\end{equation}
to a finite value :
\begin{equation} 
\frac{d}{dt}\textrm{Tr}[\rho(t)\mathcal{B}]|_{t_0+0}=\textrm{Tr}[\rho(t_0) U_{\tilde{A}}^{-1}[H_0,\mathcal{B}]U_{\tilde{A}}]\neq 0 \ ,
\end{equation}
then, $\textrm{Tr}[\rho(t)\mathcal{B}]$ being periodic and continuous, it will reach a maximum that is strictly larger within a period $T_0$. In this way, we can therefore conclude that $\rho(t_0)\notin \mathcal{S}$ and that $\mathcal{S}$ contains $\mathcal{F}$. Straigthforward calculations (see appendix \ref{appb} for details) show that a density operator $\rho\in \mathcal{S}$ if $\rho$ commutes with $\mathcal{B}$. We can then ask the conditions on $H_0$, $U_{\tilde{A}}$ and $\mathcal{O}^{(N)}$ such that these density operators be the only elements of $\mathcal{S}$ and the unique fixed points of the strategy. Appendix \ref{appb} provides a sufficient condition on these operators. However, this condition does not completely justify the control strategy and several open questions are in order : what is the domain of attraction of each fixed point, how large is the domain of the target state.... 

As a conclusion, we claim that the formulation of this control scheme for mixed-state quantum systems is general and powerful in the sense that it emphasizes the mathematical definition of a target density operator that maximizes the expectation value of the observable $\mathcal{O}$ and a systematic strategy to reach it through a set of unitary perturbations applied at determined time intervals. In particular, the construction of the target constitutes the real originality of the previous scheme in comparison with the one for pure-state quantum systems. The application of this control scheme to molecular alignment / orientation dynamics will be the subject of the next section, where the definition of the target will be highlighted and the efficiency of the preceding strategies will be shown numerically.        
\section{Control of the molecular alignment / orientation dynamics} \label{section3}
This section is devoted to the application of the strategy for controlling the alignment / orientation dynamics of polar diatomic molecules.
\subsection{Description of the model}
We consider a molecule described in a rigid-rotor approximation interacting with a linearly polarized electromagnetic pulse. The following molecular Hamiltonian $H=\varepsilon J^2-V_{a;o}(s)$, already investigated in a previous paper \cite{sugny3} dealing with the control of pure-state quantum systems, will serve as an illustration. The time-dependent Schr$\textrm{\" o}$dinger equation which governs the dynamics can be written as :
\begin{equation} \label{cont1}
i\frac{\partial}{\partial s}\psi (\theta,\phi;s)=[\varepsilon J^2-V_{a;o}(s)]\psi(\theta,\phi ;s) \ ,
\end{equation}
where $J^2$ is the angular momentum operator, $\theta$ and $\phi$ denoting respectively the polar angle and the azimuthal angle. The operators $V_{a;o}$, corresponding to the radiative interaction terms for the alignment / orientation processes, are defined respectively as follows :
\begin{equation} \label{cont2}
V_a(s)=E_a^2(s)\cos^2\theta+F_a^2(s) \ ,
\end{equation}
and 
\begin{equation} \label{cont3}
V_o(s)=E_o(s)\cos\theta \ ,
\end{equation}
$E_{a;o}$ and $F_a$ being the dimensionless interaction strength. These time-dependent functions are equal to :
\begin{eqnarray} \label{cont4}
E_a^2(s)&=& \Delta\alpha\tau f^2(\tau s)/2 \\ 
F_a^2(s)&=& \alpha_\perp \tau f^2(\tau s)/2 \\
E_o(s) &=& \mu_0 \tau f(\tau s) \ ,
\end{eqnarray}
where $\mu_0$ is the permanent dipole moment, $\Delta \alpha=\alpha_\parallel-\alpha_\perp$ is the difference between the parallel $\alpha_\parallel$ and perpendicular $\alpha_\perp$ components of the polarizability tensor and $f(t)$ is the enveloppe of the laser pulse. Note that Eq. (\ref{cont1}) is expressed in terms of the dimensionless parameter $\varepsilon=\tau B$ and $s=t/\tau$, where $B$ and $\tau$ are, respectively, the rotational constant and the pulse duration. The reader is referred to Ref. \cite{sugny3} for the explicit derivation of Eq. (\ref{cont1}). We also recall that, due to the cylindrical symmetry, the projection $m$ of the total angular momentum $j$ on the field polarization axis is a conserved quantum number. 

From the practical point of view, we are interested in short duration pulses, the duration of the pulse being compared with the molecular rotational period $T_{rot}=\pi/B$. Such kicks allow us to consider a sudden approximation and to derive a simple effective evolution operator describing the interaction with the pulse \cite{dion2,sugny2}. We have 
\begin{equation} \label{cont5}
U_a=\exp [iA_a\cos ^2\theta] \ ,
\end{equation}
where $A_a=\int_0 ^1 E_a^2(s)ds$ (the contribution of $F_a^2(s)$ has not been considered as it corresponds to a pure phase factor) for the alignment process and
\begin{equation} 
U_o=\exp [iA_o\cos \theta] \ ,
\end{equation}
with $A_o=\int_0 ^1 E_o(s)ds$ for the orientation. Using these latter propagators and the fact that $[\mathcal{O}^{(N)}_{a;o},\rho_{opt}^{(N)}]=0$, one can easily check the necessary commutation relations 
\begin{equation} \label{cont6}
[U_{a;o},\mathcal{O}^{(N)}_{a;o}]=[U_{a;o},\rho_{opt}^{(N)}]=0 \ ,
\end{equation}
$\mathcal{O}_{a;o}$ being defined as $\mathcal{O}_a=\cos^2\theta$ and $\mathcal{O}_o=\cos\theta$.

The efficiency of the alignment / orientation is characterized by a thermal average of $\mathcal{O}_{a;o}$ over the rotational levels :
\begin{equation} \label{cont7}
\langle \mathcal{O}_{a;o} \rangle (s)=\textrm{Tr}[\rho(s)\mathcal{O}_{a;o}] ,
\end{equation}
with as an initial condition
\begin{equation} \label{cont9}
\rho_0=\frac{1}{Z}\sum_{m\in\mathbb{Z}}\sum_{j\geq |m|}|j,m\rangle e^{-Bj(j+1)/k_BT}\langle j,m| \ ,
\end{equation}
where $Z=\sum_{m\in\mathbb{Z}}\sum_{j\geq m} e^{-Bj(j+1)/k_B T}$ is the partition function and the $|j,m\rangle$'s the eigenvectors of $J^2$. It is to be noted that the goal of the control of molecular alignment / orientation is not only to obtain large absolute values of $\langle \mathcal{O}_{a;o} \rangle$ but also a persistance of this effect. We will see that one can increase the duration of the alignment / orientation by choosing adequately the dimension of the molecular rotational space while keeping a high efficiency of the process. Details on this point will be given in Sec. \ref{kine}.
\subsection{Kinematical bounds and dynamical realizability} \label{kine}
We first analyze the kinematical bounds of the process. For doing so, we have to reduce the physical Hilbert space $\mathcal{H}$ to a finite subspace $\mathcal{H}^{(j_{max})}$, $j_{max}$ being the highest $j$'s for which the corresponding rotational levels are significantly populated. Simple algebra shows that the dimension $N$ of $\mathcal{H}^{(j_{max})}$ is
\begin{equation} \label{cont10}
N=(j_{max}+1)^2 \ .
\end{equation}
We recall that the density operators $\rho_{opt}^{(j_{max})}$ associated with the kinematical bounds are given by [Eq. (\ref{math6})] :
\begin{equation} \label{cont11}
\rho_{opt}^{(j_{max})}=\sum_{k=1}^N \omega_{\sigma_k}|\chi^{(k)}\rangle\langle\chi^{(k)} | \ ,
\end{equation}
the $w_k$'s being the eigenvalues of the statistical density operator [Eq. (\ref{cont9})], $\sigma$ a permutation of these eigenvalues depending on the set of density operators which are dynamically attainable and $|\chi^{(k)}\rangle$ the eigenvectors of $\mathcal{O}_{a;o}^{(N)}$. If we assume the complete controllability of the system, $\langle \mathcal{O}_{a;o}^{(N)}\rangle $ is a maximum for the permutation which associates the highest eigenvalue of $\rho^{(N)}_0$ with the one of $\mathcal{O}_{a;o}^{(N)}$, and so forth. We call this maximum, $\rho_{opt}^{(j_{max})}$ the optimal density operator. It is noted that $\rho_{opt}^{(j_{max})}$ is not unique as the eigenvalues of $\mathcal{O}_{a;o}^{(N)}$ are degenerate.\\
However, a diatomic molecule driven by a linear polarized laser pulse is a system which is not completely controllable. $m$ being a good quantum number, the system is decoupled and the Hilbert space $\mathcal{H}^{(j_{max})}$ can be written as the direct sum of the dynamically invariant subspaces $\mathcal{H}^{(j_{max})}_m$ :
\begin{equation} \label{cont12}
\mathcal{H}^{(j_{max})}=\oplus_{m=-j_{max}}^{j_{max}} \mathcal{H}^{(j_{max})}_m  \ .
\end{equation}
Moreover, as the subspaces of $\mathcal{H}^{(j_{max})}_m$ of a given parity of $j$ are not coupled by the operator $\mathcal{O}^{(N)}_a$, another subdivision has to be considered for the alignment process for which we can write
\begin{equation} \label{cont12a}
\mathcal{H}^{(j_{max})}=\oplus_{m=-j_{max}}^{j_{max}} (\mathcal{H}^{(j_{max})}_{m,ev.}\oplus \mathcal{H}^{(j_{max})}_{m,od.})  \ .
\end{equation}
The subscripts $ev.$ and $od.$ correspond respectively to the even and odd values of $j$. To simplify the notation, we omit these subscripts below when confusion is unlikely.

Let $P_m$ be the projector onto the subspace $\mathcal{H}^{(j_{max})}_m$ and the observables $\rho_m^{(N)}$ and $\mathcal{O}_{o;m}^{(N)}$ be the restrictions of $\rho^{(N)}$ and $\mathcal{O}_{o}^{(N)}$ on $\mathcal{H}^{(j_{max})}_m$. We denote by $\omega_k^{(m)}$ and $\chi_k^{(m)}$ the eigenvalues of these two operators in $\mathcal{H}^{(j_{max})}_m$. As all the subspaces are initially populated, the optimal density operator for a linear polarization $\rho_{lin}^{(j_{max})}$ is the direct sum of the optimal density operators [with this time the general definition given by Eq. (\ref{math6})] $\rho_{opt}^{(m)}$ in each subspace $\mathcal{H}^{(j_{max})}_m$ \cite{schirmer3}. $\textrm{Tr}[\mathcal{O}^{(N)}_o\rho_{lin}^{(j_{max})}]$ can be written as follows :
\begin{equation} \label{cont13}
\textrm{Tr}[\mathcal{O}_o^{(N)}\rho_{lin}^{(j_{max})}]=\sum_{m=-j_{max}}^{j_{max}}\sum_{k=1}^{j_{max}-|m|+1}\omega_k^{(m)}\chi_k^{(m)}  \ ,
\end{equation}
where the $\omega_k^{(m)}$'s and the $\chi_k^{(m)}$'s are ordered for each value of $m$. Note that a similar result taking into account the parity of $j$ can be established for the alignment.

We now examine more closely the different aspects of the controllability of these processes. We begin by analysing the individual controllability of each subsystem. Within each decoupled invariant subspaces corresponding to different values of $m$, the controllability of the subsystem is related to the dimension of the dynamical Lie algebra generated by the free Hamiltonian and the coupling operator. We consider a type of Hamiltonian which has already been investigated by S. G. Schirmer and co-workers \cite{fu,schirmer3} and it was shown that each subsystem is completely controllable together with the restriction on the parity of $j$ for the alignment. 

The density operator $\rho_{lin}^{(j_{max})}$ is dynamically realizable if all the subsystems acting on $\mathcal{H}^{(j_{max})}_m$ are completely controllable, but this condition is not sufficient \cite{turinici,schirmer5} to ensure the controllability of the whole system. Indeed, it has not been proved that all the subsystems can be simultaneously controlled by the same control pulse. Appendix \ref{appc} provides numerical and theoretical results that indicate that the whole system is not simultaneously controllable. However, a less strong notion of simultaneous controllability  can be derived by taking into account the particular symmetry of the system, i.e. the fact that its dynamics is the same in the subspaces $\mathcal{H}^{(j_{max})}_m$ and $\mathcal{H}^{(j_{max})}_{-m}$. Finally, with this latter condition on the dynamical Lie algebra (see Appendix \ref{appc} for details), it can be shown that the target states are dynamically attainable for the alignment / orientation processes.
\begin{figure} 
\includegraphics[scale=0.4]{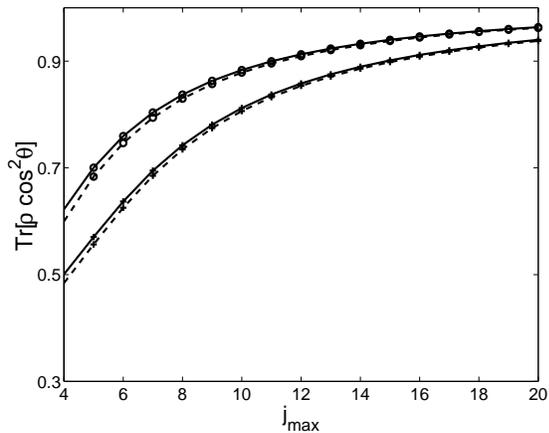}
\caption{\label{fig1} Maximum alignment efficiency as a function of $j_{max}$ (see text) for the molecule LiCl in the case $T=10$ K (crosses) and $T=5 K$ (open circles). Solid and dashed lines, which are just to guide the eye, correspond respectively to the optimal maximum and the one for a linear polarization.}
\end{figure}
\begin{figure} 
\includegraphics[scale=0.4]{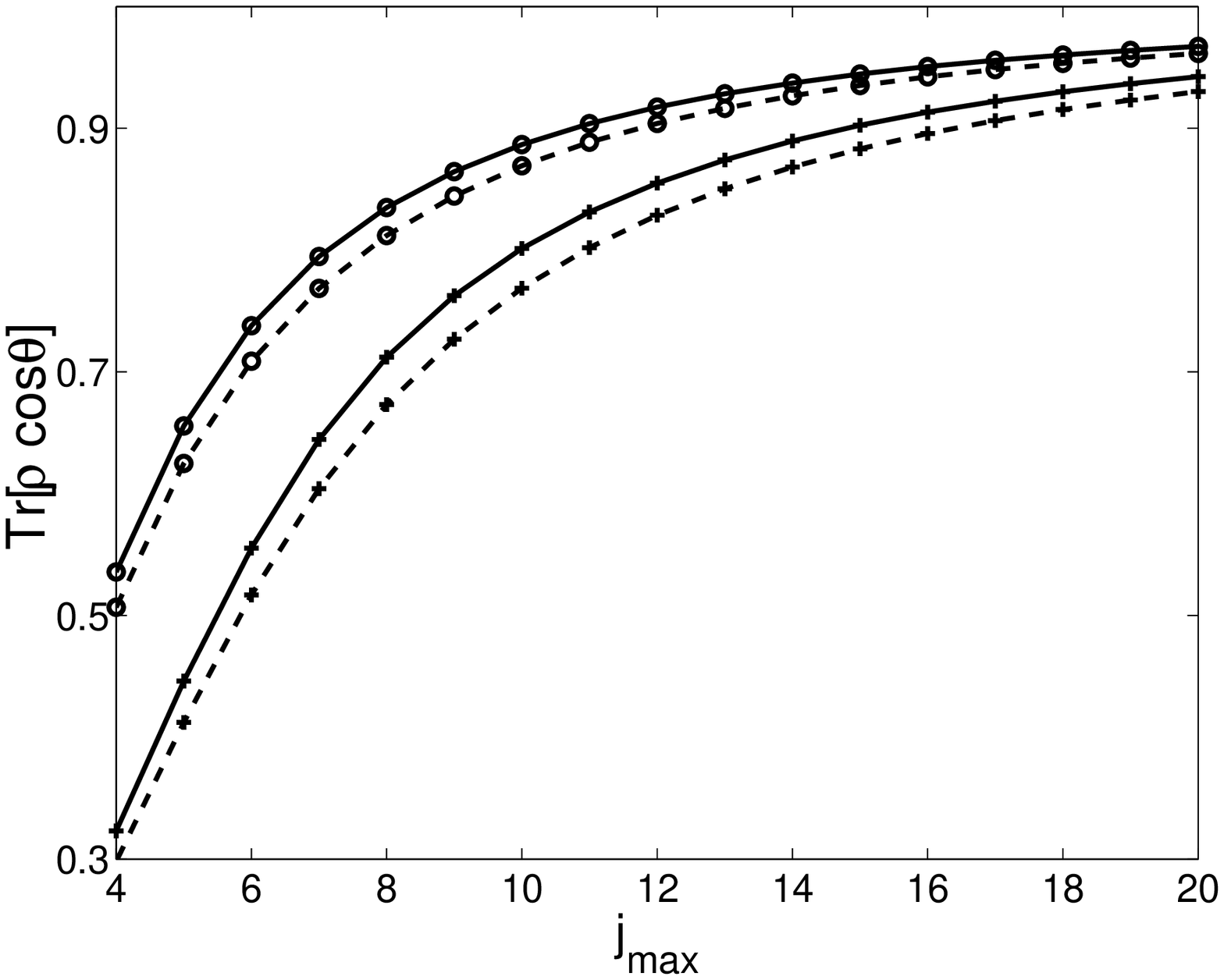}
\caption{\label{fig2} Same as Fig \ref{fig1}, but for orientation.}
\end{figure}
\begin{figure} 
\includegraphics[scale=0.4]{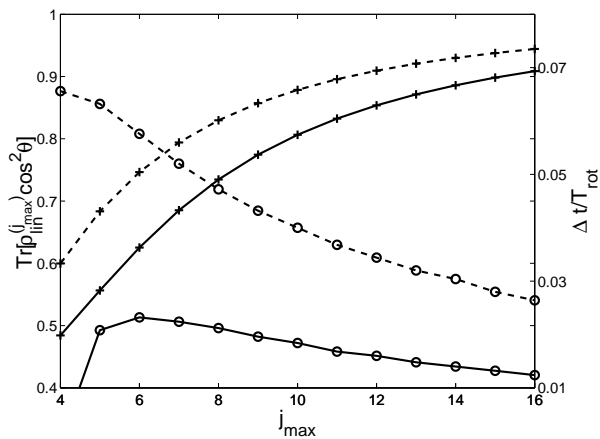}
\caption{\label{fig2a} Maximal alignment efficiencies that can be obtained with a linear polarization (crosses) and associated duration (open circles) as a function of $j_{max}$ (see text) for the molecule LiCl in the case $T=5$K (dashed lines) and $T=10$K (solid lines).}
\end{figure}
\begin{figure} 
\includegraphics[scale=0.4]{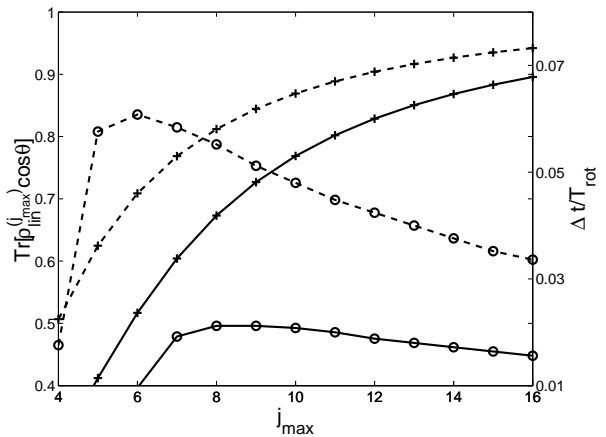}
\caption{\label{fig2b} Same as Fig \ref{fig2a}, but for orientation.}
\end{figure}

Returning to the description of the kinematical bounds, Figs. \ref{fig1} and \ref{fig2} display the maximum efficiency for the alignment / orientation processes of the molecule LiCl as a function of $j_{max}$. A very striking observation is the amount of alignment / orientation that can be achieved by the strategy using solely linear polarized pulses, namely within 10$\%$ of the optimal kinematical limit. Such an agreement is expected in the high rotational excitation limit where $\sin\theta=m/j$ tends to zero, producing alignment even though the initial set of quantum numbers $m$ has not been modified by the radiative field. Which is more unexpected is that such a circumstance occurs for low excitation (i.e. $j_{max}\simeq 8$). This leads to the important conclusion that alignment / orientation control could be satisfactorily achieved referring to linearly polarized pulses. This, as opposite to more sophisticated schemes working, for instance, with elliptically polarized pulses, may present an interest from an experimental point of view. We also notice that the optimal alignment / orientation increases as $j_{max}$ increases or the temperature $T$ decreases. However, it is important to realize that a larger $N$ corresponds to a better efficiency, but leads to a shorter duration. This point is shown in Figs. \ref{fig2a} and \ref{fig2b} where $\Delta t/T_{rot}$ is the relative duration of the field-free alignment / orientation  over which $\textrm{Tr}[\rho_{lin}^{(j_{max})}\mathcal{O}_{a;o}^{(N)}]$ remains larger than 0.5. We then see that a compromise has to be made between maximum efficiency and duration \cite{sugny}. For the alignment / orientation of the molecule LiCl, we choose $j_{max}=8$ which allows us to obtain a duration of the order of 1/20 of the rotational period and an optimal efficiency of 0.8 at least for $T=5\ \textrm{K}$. It is finally very important to note that only poor alignment / orientation efficiency and duration can be achieved for high temperatures whatever the sophistication of the excitation scheme be (different polarization directions, for instance). About 25 to $30\%$ of the efficiency is lost when heating from $5\ \textrm{K}$ to $10\ \textrm{K}$.
\subsection{Control of the dynamics}
\begin{figure} 
\includegraphics[scale=0.4]{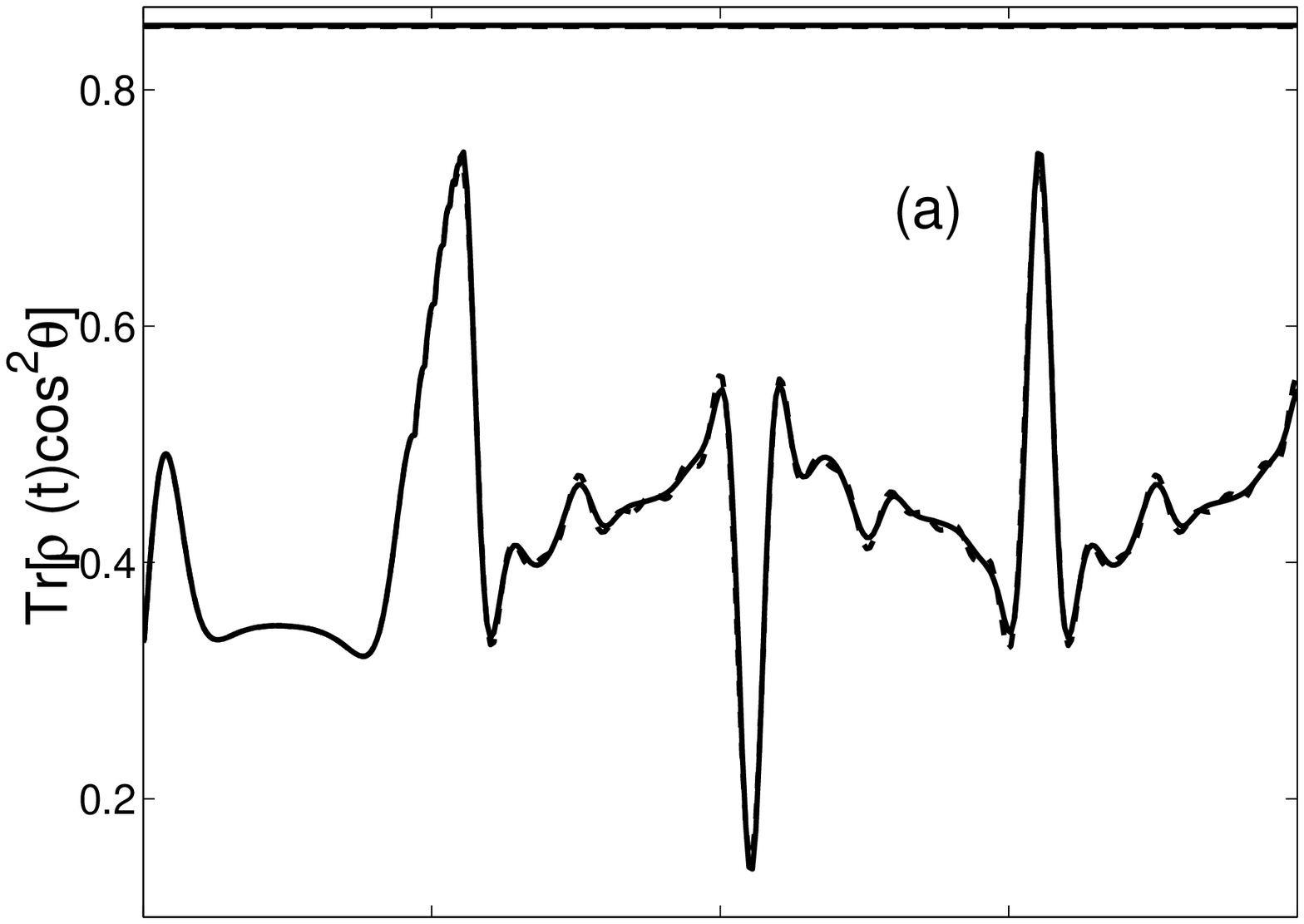}
\includegraphics[scale=0.4]{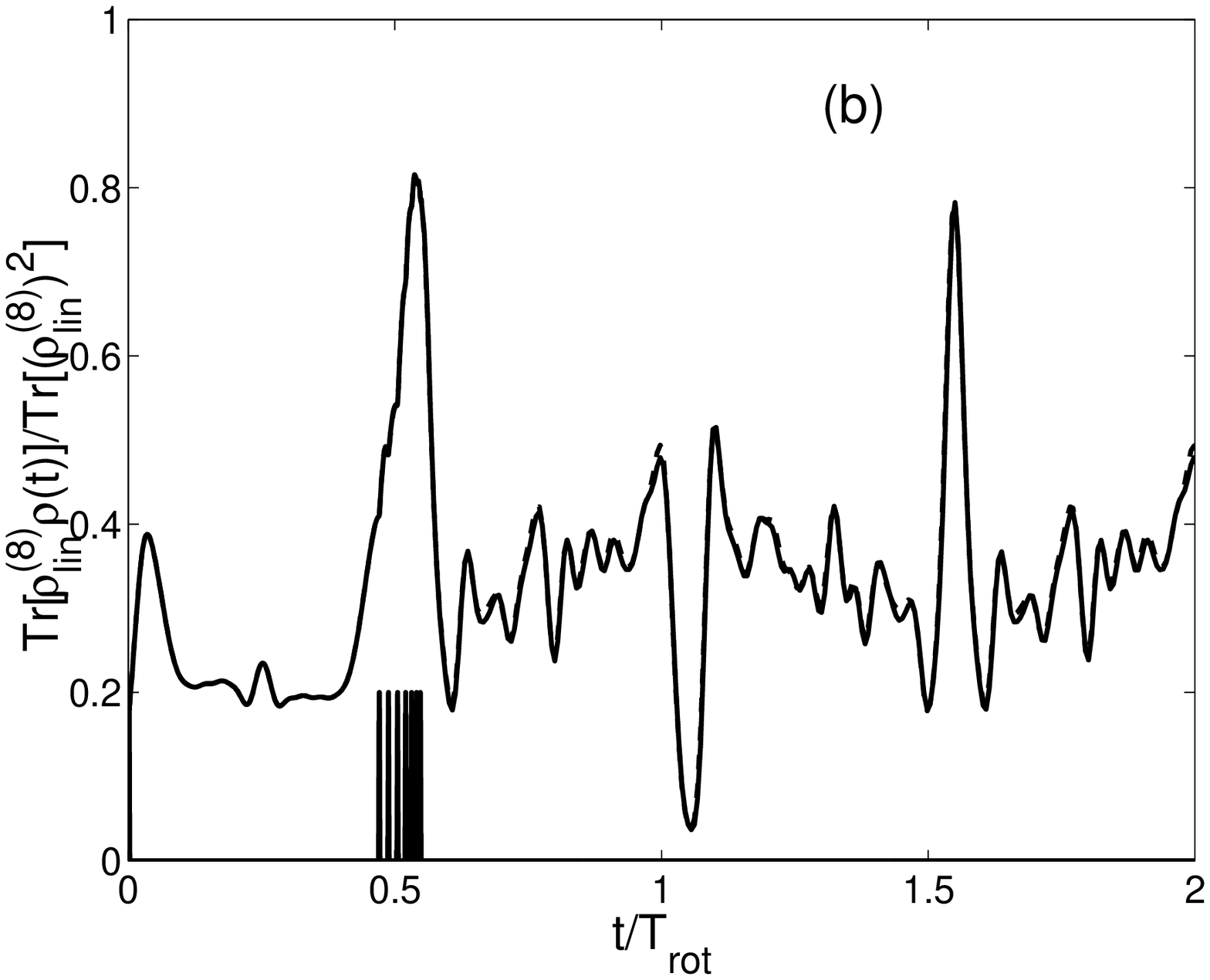}
\caption{\label{fig3} Alignment dynamics during and after the train of pulses determined by strategy $S1$, with pulses acting at the global maxima of $\textrm{Tr}[\rho(t)\cos ^2\theta]$ : panel (a) for $\textrm{Tr}[\rho(t)\cos^2\theta]$ and panel (b) for $\textrm{Tr}[\rho_{lin}^{(8)}\rho(t)]$. The solid line corresponds to the exactly propagated density operator $\rho(t)$ and the dashed line to the propagation of $\rho(t)$ in the subspace $\mathcal{H}^{(8)}$. The train of pulses is displayed on panel (b), the optimal alignment and the one for a linear polarization in $\mathcal{H}^{(8)}$ are, respectively, indicated by the horizontal solid and dashed lines on panel $(a)$.}
\end{figure}
\begin{figure} 
\includegraphics[scale=0.4]{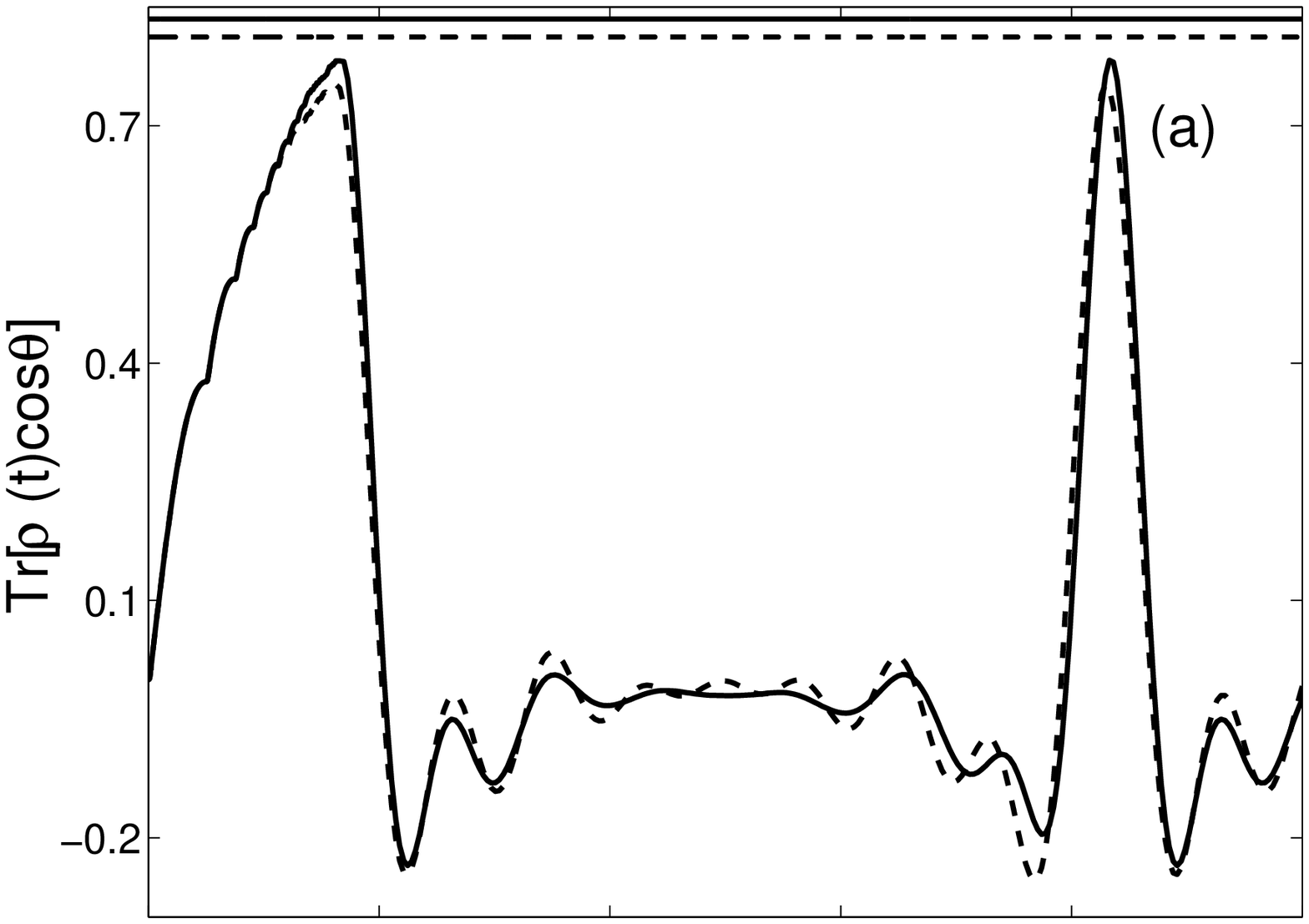}
\includegraphics[scale=0.4]{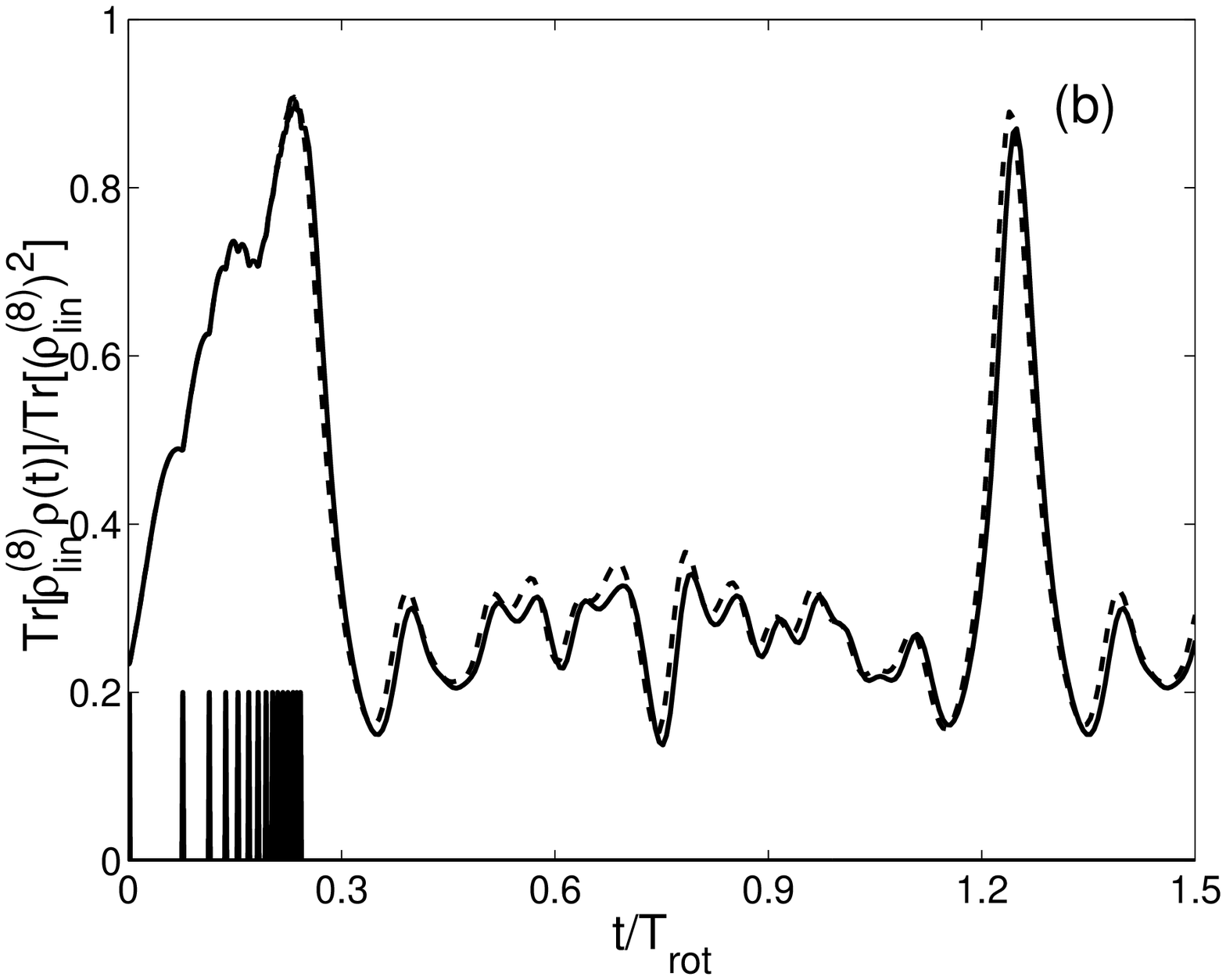}
\caption{\label{fig4} Same as Fig. \ref{fig3}, but for the orientation dynamics.}
\end{figure}
\begin{figure} 
\includegraphics[scale=0.4]{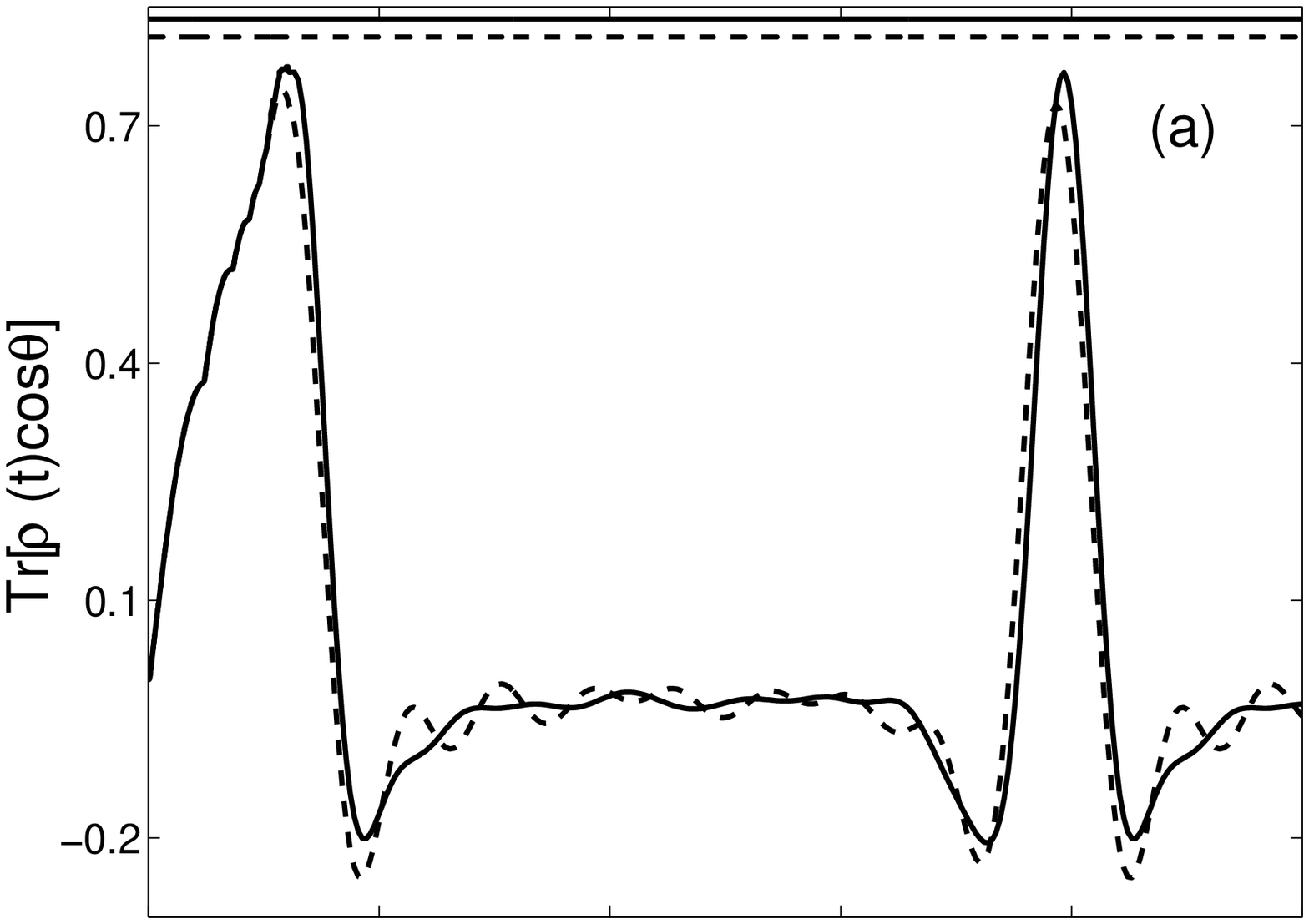}
\includegraphics[scale=0.4]{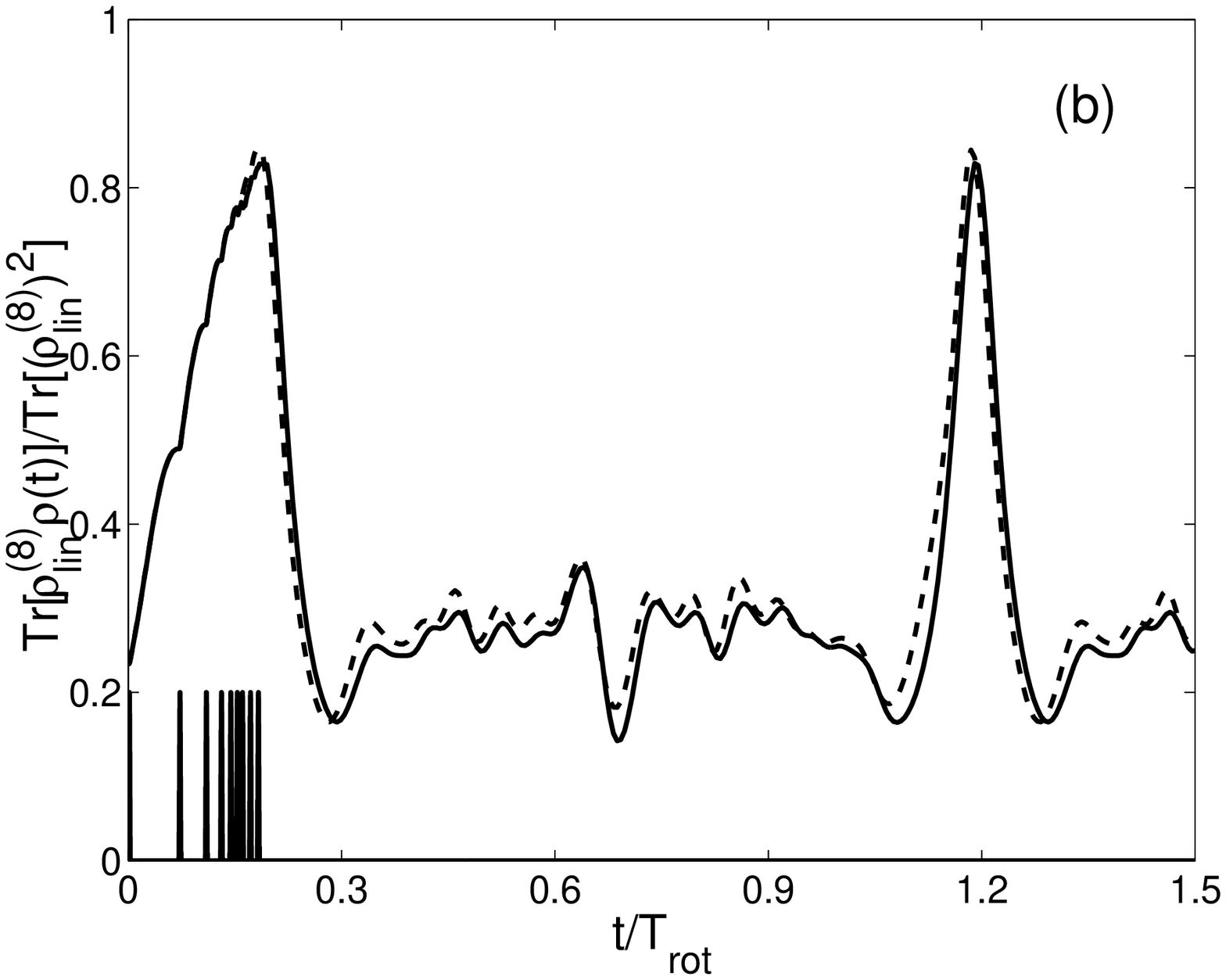}
\caption{\label{fig5} Same as Fig. \ref{fig4}, but referring to the strategy $S2$, with pulses acting at the maxima of $\textrm{Tr}[\rho_{lin}^{(8)}\rho(t)]$.}
\end{figure}
\begin{figure} 
\includegraphics[scale=0.4]{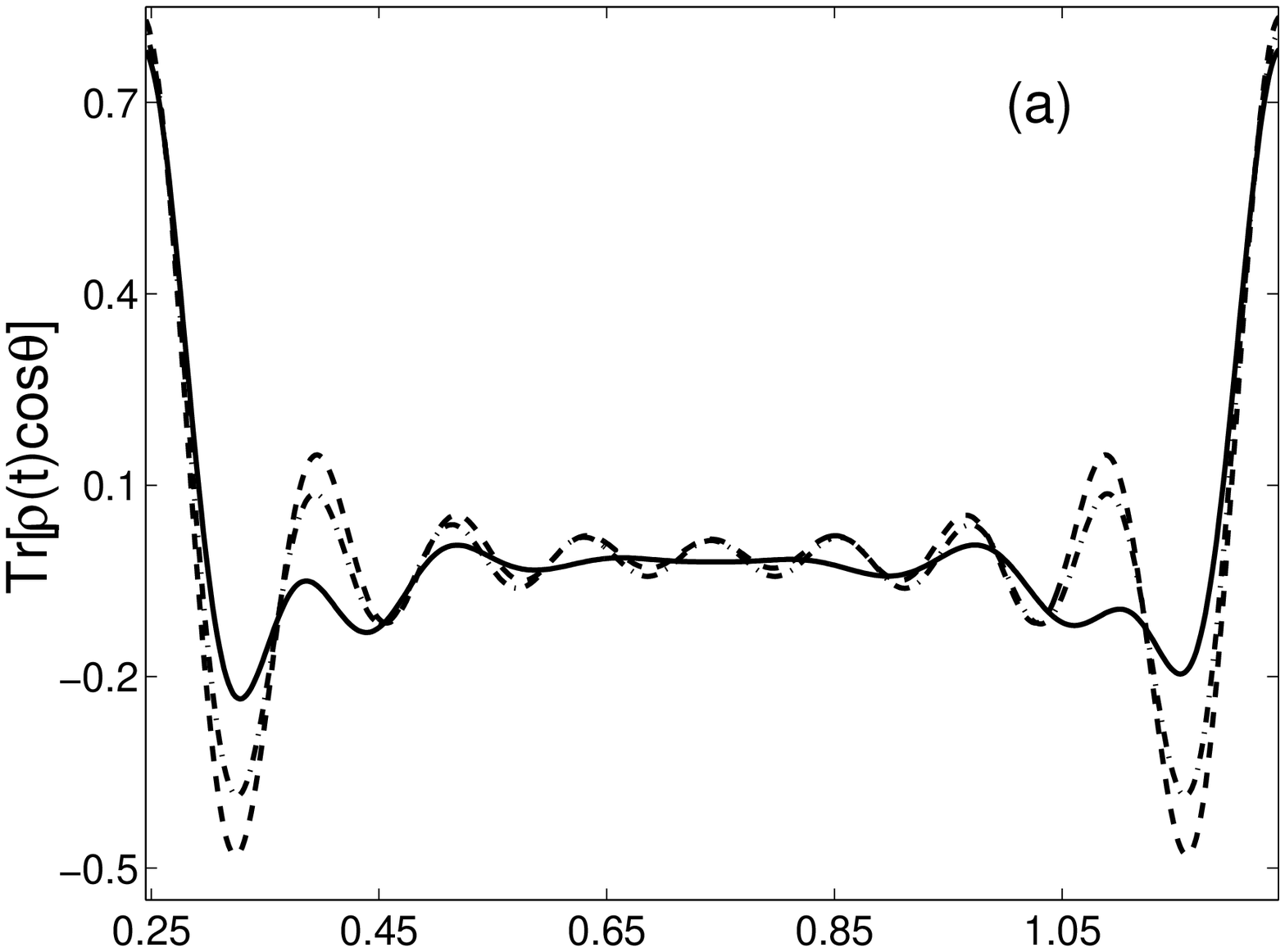}
\includegraphics[scale=0.4]{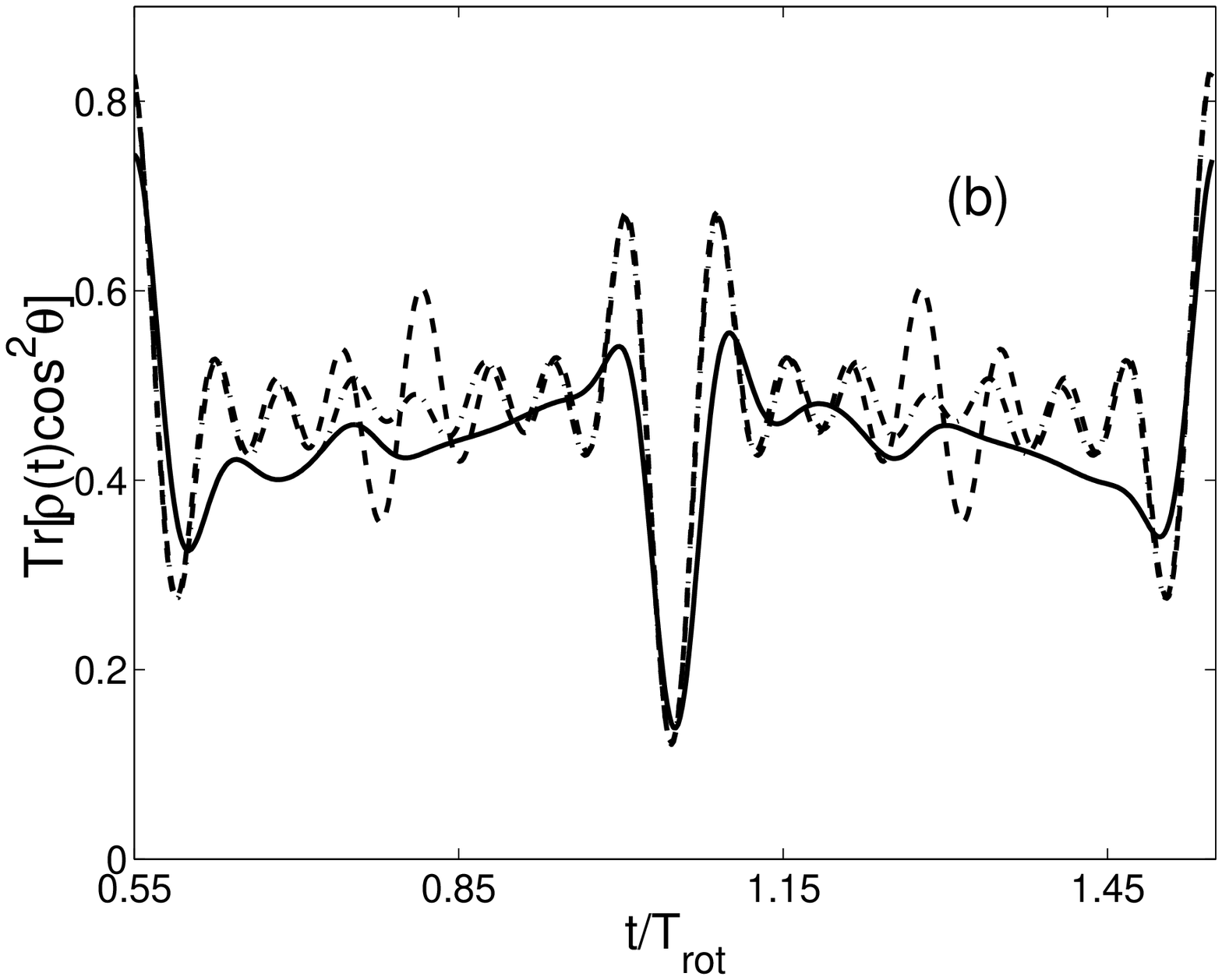}
\caption{\label{fig6} Postpulse alignment [panel (a)] and orientation [panel (b)] dynamics of the molecule LiCl after interaction with a train of short pulses (the time $t=0$ corresponds to the first kick). Solid, dashed and dash-dotted lines correspond respectively, to the averages calculated with the exact density operator, the optimal one $\rho_{opt}^{(8)}$ and the one obtained with a linear polarization $\rho_{lin}^{(8)}$. In these two cases, the strategy $S1$ is used.}
\end{figure}
The results are presented for the molecule LiCl with the rotational temperature $T=5$ K. As these results only depend on two dimensionless parameters $t/T_{rot}$ and $B/(k_BT)$, they are transferable to any particular molecule. Numerical parameters are taken to be : $A_a=2$, $A_o=2$ and $\varepsilon =0.01$. This amounts to a pulse duration of about 0.2 ps and a field amplitude of $1.5\times 10^5\ \textrm{V.cm}^{-1}$ \cite{dion2} for the case of orientation. The rotational constant $B=0.706 52\ \textrm{cm}^{-1}$ (value at the Li-Cl equilibrium distance) is chosen so as to reproduce the principal features of the diatomic molecule LiCl. In the different figures, time is indicated in fractions of the LiCl rotational period.

In this section, we are interested in the dynamical behavior of the two quantities $\textrm{Tr}[\mathcal{O}_{a;o}\rho(t)]$ and $\textrm{Tr}[\rho_{lin}^{(8)}\rho(t)]/\textrm{Tr}[(\rho_{lin}^{(8)})^2]$, which correspond respectively to the expectation value of the observable $\mathcal{O}_{a;o}$ and to the projection on the target state $\rho_{lin}^{(8)}$. Except for Figs. \ref{fig6}, it is noted that all results are displayed for times starting at the first pulse (taken thus at the origin) and extending up to one rotational period $T_{rot}$ after the last pulse. This helps in showing the complete field-free behavior of the dynamics and the well-known revival structures.
 
The results of the strategy $S1$ for the alignment and orientation processes are illustrated in Figs. \ref{fig3} and \ref{fig4}. As could be expected, the alignment dynamics presents a more strongly oscillating structure as compared with the orientation one. This difference explains the different choice of parameters $A$, $N$ and the number of kicks in these two processes. In all cases, panels (b) show that the density operator $\rho(t)$ gets close to the target state $\rho_{lin}^{(8)}$, which also shows that the rotational dynamics roughly resides within $\mathcal{H}^{(8)}$. It can furthermore be checked that the postpulse dynamics lead to very good results in terms of efficiency and duration. Indeed, an efficiency of about 0.75 and a duration of about 1/20 of the rotational period (that is about 1 ps for a light molecule like LiCl) is obtained for the orientation. To our knowledge, this is the largest duration and efficiency achieved up to date for such a thermal ensemble.

In Figs. \ref{fig5}, we display the mean values $\textrm{Tr}[\mathcal{O}_{a;o}\rho(t)]$ and $\textrm{Tr}[\rho_{lin}^{(8)}\rho(t)]/\textrm{Tr}[(\rho_{lin}^{(8)})^2]$ computed with the strategy $S2$ for the case of orientation. It is seen that very close results are obtained with respect to Figs. \ref{fig4} and the strategy $S1$, which shows the similarity between the two control schemes. Note that a smaller number of pulses, i.e. 9, is used for $S2$. 

Finally, the evolution after the last kick for the alignment / orientation processes  are plotted in Figs. \ref{fig6}, which, in addition, also display the dynamical behavior of the two optimals, $\rho_{lin}^{(8)}$ and $\rho_{opt}^{(8)}$. We notice that the control strategy allows us to obtain a density operator with free dynamics very close to the optimal one, both in efficiency and duration. Once again, these results stress the importance of the target states since they give a clear insight into the free dynamics of the molecule. 
\section{Conclusion} \label{section4}
A precise mathematical definition of a target provides a completely new insight to the control issue of a system using a time dependent external field. Complementary to what is done for pure-state quantum systems \cite{sugny3} where the target is an eigenvector of a given observable projected onto a finite dimensional Hilbert subspace, the target in the mixed-state case is a non-trivial density matrix constructed as a specific combination of the eigenvectors of a given observable through weighting factors related with the initial statistics of the mixed system. Once the target is identified and its attainability established, two possible strategies to reach it are thoroughly analyzed. They amount to apply a series of unitary perturbations each time $\textrm{Tr}[\rho(t)\mathcal{O}^{(N)}]$ or $\textrm{Tr}[\rho(t)\rho_{opt}]$ reaches a global maximum (or minimum) under free evolution.

An illustration is given by considering the laser control of molecular alignment / orientation in thermal equilibrium. Although some previous works have already used the strategy of applying trains of short laser pulses, the originality resides here in the practical construction of the target density matrix and in comparing the optimal alignment / orientation actually obtained to the ones which are the best theoretically possible within the reduced subspace supporting the dynamics.

As a prospect, an important open question in this field is the applicability of the present method to more complicated systems involving, for instance, an interaction with a physical environment or a dissipative system \cite{shlomo,tannor}. The answer is not obvious because the strategy would involve non-unitary evolution and special attention has to be paid to decoherence \cite{beige,lidar}. Moreover, for an important class of dissipative systems such as laser cooling, the dissipation can increase the purity of the state \cite{shlomo,tannor,schirmer4}. In this case, the optimal target state would be the one defined in Ref. \cite{sugny} for pure-state quantum systems.
\begin{acknowledgments}
The authors thank G. Turinici for helpful discussions.
\end{acknowledgments}
\appendix

\section{Analysis of the set of fixed points} \label{appb}
In this appendix, we analyse the set $\mathcal{S}$ with the hypothesis of complete controllability. For that purpose, we introduce the vector space $\mathcal{V}$ generated by the operators $[H_0,\mathcal{B}]$ and $U_{\tilde{A}}^{-1}[H_0,\mathcal{B}]U_{\tilde{A}}$, where $\tilde{A}\in \mathbb{R}$. One can reformulate Eqs. (\ref{math11}) and (\ref{math12}) by stating that $\rho$ is an element of $\mathcal{S}$ if $\rho \in \mathcal{V}^\perp$, where $\mathcal{V}^\perp$ is the orthogonal space of $\mathcal{V}$. It is noted that $\mathcal{V}^\perp$ does not only contain density operators. Moreover, it can also be shown that $\rho\in\mathcal{S}$ if the density operator $\rho$ commutes with $\mathcal{B}$. Indeed, as $[\mathcal{B},U_{\tilde{A}}]=0$, this remark can be easily checked by calculating the following expression :
\begin{eqnarray} \label{math15}
\textrm{Tr}[\rho U_{\tilde{A}}^{-1}[H_0,\mathcal{B}]U_{\tilde{A}}]&=&\textrm{Tr}[\rho [U_{\tilde{A}}^{-1}H_0U_{\tilde{A}},\mathcal{B}]] \\
&=& \textrm{Tr}[(\mathcal{B}\rho-\rho\mathcal{B}) U_{\tilde{A}}^{-1}H_0U_{\tilde{A}}] \ .
\end{eqnarray}
At this point, the question that naturally arises is the condition on $H_0$, $\mathcal{B}$ and $U_{\tilde{A}}$ such that $\mathcal{S}$ only contains these latter elements. The condition is the following : The density operators which commute with $\mathcal{B}$ are the only elements of $\mathcal{S}$ if and only if the dimension of $\mathcal{V}$ is $N^2-\sum_i n_i^2$, where the $n_i$'s $(i=1,\cdots ,P)$ are the multiplicities of the eigenvalues of $\mathcal{B}$.\\
The proof is straightforward. We note $|b_{i;k}\rangle (k=1,\cdots ,n_i)$ the eigenvectors of $\mathcal{B}$. A density operator $\rho^{(N)}$ such that $[\rho^{(N)},\mathcal{B}]=0$ is of the form :
\begin{equation} \label{math16}
\rho^{(N)}=\sum_{i=1}^P\sum_{k,k'=1}^{n_i} \alpha_{i;k;k'} |b_{i;k}\rangle\langle b_{i;k'}| \ ,
\end{equation}
where the $\alpha_{i;k;k'}$'s are complex numbers such that the eigenvalues $\omega_k$ of $\rho^{(N)}$ fulfill the two conditions : $0\leq \omega_k\leq 1$ and $\sum_{k=1}^N \omega_k=1$. The vector space generated by these operators is of dimension $\sum_i n_i^2$ and it is a subspace of $\mathcal{V}^\perp$. In addition, $U_{\tilde{A}}[H_0,\mathcal{B}]U_{\tilde{A}}^{-1}$ being a skew-hermitian operator, it can readily be shown that $\mathcal{L}$ contains $\mathcal{V}$. One finally deduces that the dimension of $\mathcal{V}$ is at most $N^2$ (the complete controllability being assumed), which completes the proof (we recall that $\dim\mathcal{V}+\dim\mathcal{V}^\perp=N^2$).
\section{Simultaneous mixed-state controllability of the molecular alignment / orientation dynamics} \label{appc}
In this appendix, we consider the simultaneous controllability of the alignment / orientation  processes driven by a linear laser field. A sufficient and necessary condition to ensure the simultaneous mixed-state controllability of a system is given by a condition on the dimension of the dynamical Lie algebra $\mathcal{L}$ (generated by the skew-hermitian operators $i(H_0)^{(N)}$ and $i\mathcal{O}_{a,o}^{(N)}$) which must be equal to \cite{schirmer5} 
\begin{equation} \label{appc1}
\textrm{D}=r+(J_{max}+1)^2-1+2\sum_{m=1}^{J_{max}}[(J_{max}-m+1)^2-1] \ ,
\end{equation}
where $r$ is the rank of the matrix 
\begin{eqnarray} \label{appc11}
T= \left( \begin{array}{cc}
\textrm{Tr}[H_0]_{m=j_{max}} & \textrm{Tr}[\mathcal{O}_{a;o}]_{m=j_{max}} \\
\vdots  & \vdots  \\
\textrm{Tr}[H_0]_m & \textrm{Tr}[\mathcal{O}_{a;o}]_m \\
\vdots  & \vdots  \\
\textrm{Tr}[H_0]_{m=-j_{max}} & \textrm{Tr}[\mathcal{O}_{a;o}]_{m=-j_{max}}  \end{array} \right) \ ,
\end{eqnarray}
For the case of orientation, straightforward calculations lead to the following expression :
\begin{eqnarray} \label{appc12}
T= \left( \begin{array}{cc}
j_{max}(j_{max}+1) & 0 \\
\vdots  & \vdots  \\
\sum_{k=|m|}^{j_{max}}k(k+1) & 0 \\
\vdots  & \vdots  \\
j_{max}(j_{max}+1) & 0  \end{array} \right) \ ,
\end{eqnarray}
which implies that the rank $r$ of this matrix is $r=1$. In the case of alignment where the trace has to be taken separately in the subspace $\mathcal{H}^{(j_{max})}_{m}$ for the odd and even values of $j$, this rank is equal to $r=2$.

We now analyse the simultaneous controllability of the orientation process. We have already mentionned that each subsystem for each value of $m$ is completely controllable. We consider independently the two subsystems corresponding to the values $m=\pm(j_{max}-1)$, $j_{max}>1$, which can be viewed as two non-interacting two-level systems. In this case, we have :
\begin{equation} \label{appc2}
(\mathcal{O}_o)_S=d\sigma_x\oplus d\sigma_x \ ,
\end{equation}
and
\begin{equation} \label{appc3}
(H_0)_S=\textrm{diag}(E_0^{(-)},E_1^{(-)})\oplus \textrm{diag}(E_0^{(+)},E_1^{(+)}) \ ,
\end{equation}
where $d=1/\sqrt{1+2j_{max}}$, $E_0^{(-)}=E_0^{(+)}=(j_{max}-1)j_{max}$ and  $E_1^{(-)}=E_1^{(+)}=(j_{max}+1)j_{max}$. The subscript $S$ indicates the restriction of the operator to the two subsystems. As $E_1^{(+)}-E_0^{(+)}=E_1^{(-)}-E_0^{(-)}$, it can be shown \cite{schirmer5} that this system and therefore the whole system (when $j_{max}>1$) are not simultaneously controllable. Note that such a result can also be established for the alignment process.

At this point, it is important to remark that a constraint on the dynamics has not been taken into account in the condition of Eq. (\ref{appc1}). Using the fact that the dynamics in the subspaces $\mathcal{H}^{(j_{max})}_{m}$ and $\mathcal{H}^{(j_{max})}_{-m}$ is the same, a new criterion on the dimension of the dynamical Lie algebra can be derived for the orientation process, where only the subspaces $\mathcal{H}^{(j_{max})}_{m}$ with $m\geq 0$ are considered :
\begin{equation} \label{appc4}
\textrm{D'}=r+(J_{max}+1)^2-1+\sum_{m=1}^{J_{max}}[(J_{max}-m+1)^2-1] \ .
\end{equation}
A similar condition can be established for the alignment together with the condition on the parity of $j$.

From a numerical point view, we can calculate the dimension of $\mathcal{L}$ for low values of $j_{max}$ \cite{schirmer2}. The results of these computations are presented in Tab. \ref{tab1} and \ref{tab2}.
\begin{table}[ht]
  \caption{\label{tab1} Dimension of the dynamical Lie algebra $\mathcal{L}$
    as a function of $j_{max}$ for the orientation
    dynamics. The third and fourth columns indicate respectively the dimensions D and D' of $\mathcal{L}$ needed for the general [Eq. (\ref{appc1})] and restricted [Eq. (\ref{appc4})] simultaneous controllabilities of the system.} 
  \begin{center} 
\begin{tabular}{c|c|c|c} 
\hline
$J_{max}$ & $\dim(\mathcal{L}$) & D & D' \\
\hline
\hline
1 & 4 & 4 & 4\\
2 & 12 & 15 & 12\\
3 & 27 & 38 & 27\\
\hline 
\end{tabular}
\end{center}
\end{table}

\begin{table}[ht]
  \caption{\label{tab2} Same as Tab. \ref{tab1}, but for alignment.} 
  \begin{center} 
\begin{tabular}{c|c|c|c} 
\hline
$j_{max}$ & $\dim(\mathcal{L}$) & D & D'  \\
\hline
\hline
1 & 2 & 5 & 2\\
2 & 5 & 16 & 5\\
3 & 11 & 39 & 11\\
\hline 
\end{tabular}
\end{center}
\end{table}
The results of Tabs. \ref{tab1} and \ref{tab2} suggest the hypothesis that the alignment / orientation processes are simultaneously controllable in the restricted sense defined above. Moreover, as by construction the target states fulfill the constraints of the dynamics, we can conclude that these targets are dynamically attainable.

\end{document}